# Acoustic resonance by fish schools:
# A proposal for the schooling mechanism

F. Meseguer, and F. Ramiro-Manzano

*Instituto de Tecnología Química (CSIC-UPV).*

*Universitat Politècnica de València, Av. Tarongers s/n 46022, Valencia, Spain*

*Email: fmese@fis.upv.es*

**Abstract**

*The acoustic properties of a fish school have been modelled using a cloud of bubbles. Similar to how bubble clouds present a collective monopole mode, a fish school also shows a collective breathing mode in which the whole school resonates as a single body. Here, we conjecture that the underwater acoustic noise of the ocean amplified by the multiple scattering of swim bladders in the fish school might produce attractive acoustic forces that are strong enough to account for the schooling mechanism. Our model predicts the presence of attractive Bjerknes forces, as large as 30% of the fish weight for 20-metre large fish schools, and that the attractive force on every fish in the centre of the school is cancelled when the fish increase/decrease the volume/pressure of its gas bladder.*

*To the best of our knowledge, this is the first example of a purely mechanical force that might contribute to fish being bound to or released from the school. The study may lead to new areas of research in many scientific fields beyond the nearest-neighbour interaction mechanism customarily used, and it would help in our understanding of collective processes of systems in ecology, physics, chemistry, and biology.*

Many scientists have been puzzled by the beautiful arrangements of fish usually called fish schools[1]. The pioneering work of Nicolis and Prigogine[2] on self-organization was followed by many others trying to understand this collective behaviour in terms of the interactions between individuals forming the group[3-5]. Faucher et al.[6] showed that the lateral line organ (LLO)[7] plays an important role in the schooling mechanism. All reported fish schooling models have been based on the fish active response to an external stimulus[3-7]. Here, we show that a fish inside a school is passively bound to its position. The presence of the underwater acoustic noise of the ocean[8], amplified by multiple scattering of the swim bladders in the school, might produce strong attractive acoustic Bjerknes[9] forces comparable to the weight of the fish.

The fish swim bladder is strongly affected by sound waves, and this bladder has been modelled by a gas bubble in water[10] using the well-known Minnaert's[11] monopole expression,

$$\omega_0 = R_0^{-1}\sqrt{3\gamma p_0/\rho} \quad (1)$$

where $R_0$ is the bubble radius, $\rho$ is the water density, $\gamma$ tis the specific heat ratio for the gas in the bubble, and $p_0$ is the static pressure of the gas. There has being increasing interest in extending this bubble model to study the acoustical properties of fish schools[12-17]. In the pioneering paper of Feuillade et al.[12], a fish school is emulated by a three-dimensional arrangement of bubbles, called a bubble cloud. This work uses the results of Otma[14] and d'Agostino and Brennen[15] for the acoustical properties of a bubble cloud. It was shown that the resonant mode with the lowest frequency value (monopole resonance) of a cloud is similar to the fundamental breathing mode of a single large bubble with a total volume equivalent to the volume of all bubbles forming the cloud. That is, the bubble resonates at a frequency value, $\nu_1$, given by the following expression (Figure 1):

$$\nu_1 \cong \nu_0 / \left( \left( R_1/R_0 \right) \alpha^{1/2} \right) \quad (2)$$

where $R_1$ and $R_0$ are the radii of the cloud and the bubble, respectively, and $\alpha$ is the filling fraction of the bubble cloud. Although the expression (2) is a simplified model, it provides useful information regarding the low-frequency value of the fundamental collective breathing mode (see Supplementary Information).

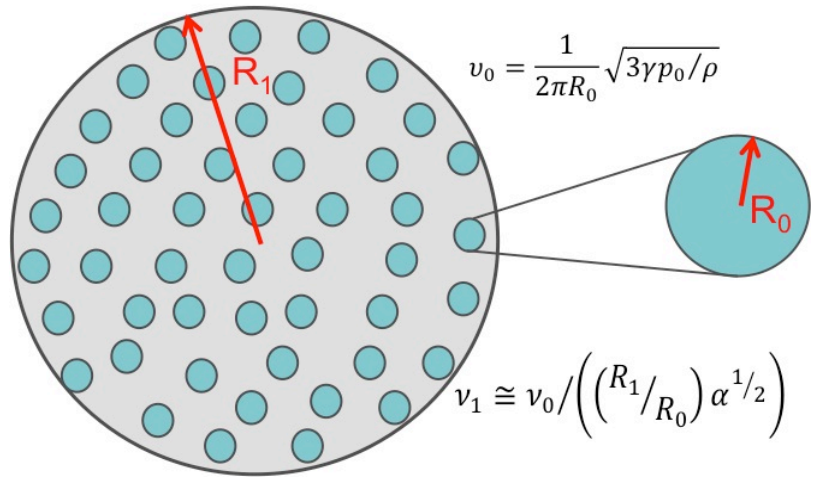

*Figure 1. Cloud bubble model with radius $R_1$ and filling fraction $\alpha$, used for understanding the acoustical properties of a fish school. The building block corresponds to a single bubble mimicking a single fish with a spherical swim bladder of radius $R_0$.*

Many researchers have studied Bjerknes forces[18-22]. The secondary Bjerknes force, $F_{SB}(\alpha, \beta)$, acts between two bubbles, $\alpha$ and $\beta$ with radii $R_{\alpha o}$ and $R_{\beta o}$ respectively, with a distance between bubbles of $L_{\alpha\beta}$. The bubbles resonate at angular frequency values $\omega_\alpha$ and $\omega_\beta$. When subjected to an acoustic field with a complex amplitude of pressure $A_m$ and an angular frequency $\omega$, the force can be written as[18]

$$\boldsymbol{F}_{SB}(\omega,\alpha,\beta) = \frac{2\pi |A_m|^2 R_{\alpha o} R_{\beta o}}{\rho_0 \omega^2 L_{\alpha\beta}{}^2 |D_{\alpha\beta}|^2} \left[ \left( \frac{\omega_\alpha^2}{\omega^2} - 1 + \frac{R_{\alpha o}}{L_{\alpha\beta}} \right) \left( \frac{\omega_\beta^2}{\omega^2} - 1 + \frac{R_{\beta o}}{L_{\alpha\beta}} \right) + \delta_\alpha \delta_\beta \right] \boldsymbol{u}_{\alpha\beta} \quad (3)$$

where

$$D_{\alpha\beta} = \left[ \left( \frac{\omega_\alpha^2}{\omega^2} - 1 - i\delta_\alpha \right) \left( \frac{\omega_\beta^2}{\omega^2} - 1 - i\delta_\beta \right) - \frac{R_{\alpha o} R_{\beta o}}{L_{\alpha\beta}{}^2} \right] \quad (4)$$

Here, $\rho_0$ is the density of the fluid (water in this case), and $\delta_\alpha$ and $\delta_\beta$ are the damping losses of the α, β bubbles that depend on both the bubble size and the frequency value. $F_{SB}(\omega,\alpha,\beta)$ is a vector along the $u_{\alpha\beta}$ direction. In general, the damping loss δ has several contributions[13,18], and in the case of fish and cm size bubbles (see Supplementary Information), we can write this loss as[16,17]

$$\delta \sim \delta_{rad} + \delta_{vis}$$

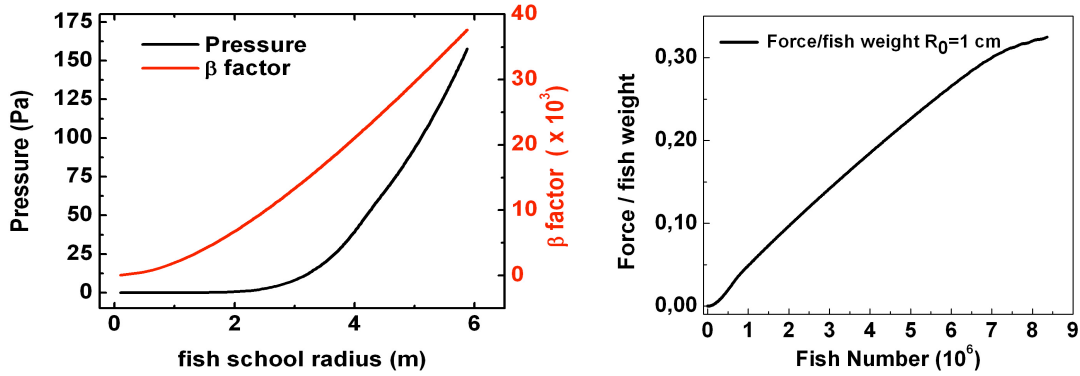

*Figure 2. Bjerknes force and pressure. Left panel: Total pressure (black curve) and β factor (red curve) as a function of the bubble cloud radius at a sea depth h=10 m. Right panel: Total force per fish weight as a function of the number of fish in the fish school at a sea depth h=20 m (see text).*

Secondary Bjerknes forces are attractive when $\omega \lessgtr \omega_\alpha, \omega_\beta$, and repulsive when $\omega_\alpha > \omega > \omega_\beta$. As the fish school is formed by fish that are similar in size, we can assume all bubbles have the same size, $R_0$, and the same gas pressure, $p_0$. Thus, $R_{\alpha 0} = R_{\beta 0} = R_0$; $\omega_\alpha = \omega_\beta = \omega_1$; and $\delta_\alpha = \delta_\beta = \delta_1$. We assume all bubbles are located at different points on a Body Centred Cubic (BCC) lattice, with Cartesian coordinates α=(ijk) and β=(lmn). It is very difficult to calculate the Bjerknes force produced by the background noise of the ocean (BNO) between a single bubble and the rest of the bubble cloud, including multiple scattering effects, even for a modest number of bubbles. Therefore, we have made the following assumptions:

A) The force on a single bubble (lmn) is the total contribution of the rest of the bubbles (ijk) integrated over the frequency region of interest. We assume a resonant frequency $\omega_1$ for all bubbles of radius $R_0$. Then, as we have a single resonant frequency, *all forces become attractive*.

B) We also know that a cloud of bubbles behaves as an acoustic cavity, where the sound resonates with a lifetime value depending on the quality factor Q of the cavity, defined as $Q = 1/\delta_1$. Therefore, the (lmn) bubble feels the scattering from the (ijk) bubble for a period of time $\tau = Q/\nu_1$, which, in the case of large fish schools (see the Supplementary Table 1), can reach values of several seconds. Therefore, the Bjerknes force between each bubble pair should be reinforced by a factor, $\beta_{ijklmn}$, defined as $\beta_{ijklmn} = \tau c / L_{ijklmn}$, where c is the sound velocity in water and $L_{ijklmn}$ is the distance between the bubbles located at (ijk) and (lmn).

The expression of the Bjerknes force for multiple scattering (MS) would be as follows (see Supplementary Information):

$$F_{SB}^{TOTAL\ MS}(lmn) = \sum_{\forall(ijk \neq lmn)} \int_{\Omega_1}^{\Omega_2} F_{SB}(\omega, ijk, lmn)\beta_{ijklmn}(\omega)d\omega \quad (5)$$

$$F_{SB}(\omega, ijk, lmn) = \frac{2\pi|A_m|^2 R_0^2}{\varrho_0 \omega^2 L_{ijklmn}{}^2 |D_{ijklmn}|^2}\left[\left(\frac{\omega_1^2}{\omega^2} - 1 + \frac{R_O}{L_{ijklm}}\right)^2 + \delta_1^2\right]u_{ijklmn} \quad (6)$$

$$|D_{ijklmn}|^2 = \left[\left(\left(\frac{\omega_1^2}{\omega^2} - 1\right)^2 + \delta_1^2\right)^2 - \frac{2R_0^2}{L_{ijklmn}{}^2}\left(\left(\frac{\omega_1^2}{\omega^2} - 1\right)^2 - \delta_1^2\right) + \frac{R_0^4}{L_{ijklmn}{}^4}\right] \quad (7)$$

where

$$L^2{}_{ijklmn} = \frac{d^2}{3}[(i-l)^2 + (j-m)^2 + (k-n)^2] = \frac{d^2}{3}[ijklmn]^2$$

When mean field theory comes into play, the bubble cloud (and also the fish school) would behave as a large bubble of effective radius $R_{01}$[14,15] (on the order of magnitude of several tens of centimetres). It is important to emphasize that both the resonant frequency value $v_1$ and the damping constant $\delta_1$ decrease dramatically for increasing fish school sizes. This fact has two main consequences (see the Supplementary Information):

1. The resonant wavelength value is much larger than the fish school size regardless of the school size; i.e., all fish in the school would feel each other.
2. The amplification factor, $\beta_{ijklmn}$, can reach very large values; thus, it might dramatically amplify the Bjerknes forces.

From the fish school example with $R_1$=5 metres (see fourth row in the Supplementary Table 1), we obtain very long dwelling time values, about $\tau$=2 seconds, which correspond to very large amplification factors, around $\beta=10^4$.

The expression (5) is difficult to handle since it involves calculating the forces from all fish. The first task is to locate all neighbours (up to eight million) in a very large BCC lattice[23]. It also involves calculating the components of the force. However, as we are considering a spherical fish school arranged in a BCC lattice, we can safely assume an isotropic force on a single fish located in the centre of the fish school. Then, the pressure $P_{SB}^{TOTAL-MS}(000)$ acting on the gas bladder can be written as

$$P_{SB}^{TOTAL-MS}(000) = \frac{1}{4\pi R_0^2}F_{SB}^{TOTAL-MS}(000)$$
$$= \frac{1}{4\pi R_0^2}\sum_{\forall(ijk \neq 000)} \int_{\Omega_1}^{\Omega_2} F_{SB}(\omega, ijk, 000)\beta_{ijk000}(\omega)d\omega \quad (8)$$

That is, we assumed that the total pressure can be written as the sum of the module of all forces equally distributed over the surface of the gas bladder. We numerically calculated both the pressure $P_{SB}^{TOTAL-MS}(000)$ and the total force $F_{SB}^{TOTAL-MS}(000)$ for fish schools of different sizes at two different sea depths, h=10 m and h=20 m, as shown in the Supplementary Table 1. We considered the total force over the BNO in the frequency range 1 Hz < ν < 1000 Hz[8] (see the Supplementary Figure 1). Figure 2 (left panel) shows the evolution of total pressure for different values of fish school radius. It also shows the evolution of the β factor (β=τc/d) as a function of the size of the fish school. In all cases, we used the following parameter values: a bubble radius $R_0=10^{-2}$ metres, and a filling fraction factor α=0.5%. This model represents fish schools of different sizes composed of fish of length L=29 cm, periodically distributed in BCC lattice, with a nearest-neighbour distance (NND) between fish of d=0,1 metres. We used a realistic value for the shear viscosity[16], μ=50 N s/m², which is associated with the swim bladder. We followed the Love model for calculating the fish volume, and we assumed that the gas bladder represents 4% of the fish volume[17]. The β=τc/d factor is the maximum amplification factor due to multiple scattering effects. The acoustic pressure and the β factor increase nonlinearly with the fish school radius. The right panel in Figure 2 shows the evolution of the ratio of the total force to the fish weight as a function of the number of fish in the school at a sea depth of h=20 m. The total force increases quasi-linearly with the number of fish in the school, although it shows a certain tendency to saturation. This unexpected behaviour can be explained due to the very large values of both the multiple scattering factor β (Figure 2) and the large number of neighbours (see the Supplementary Figure 2). The calculated total forces can reach values of around 30% of the fish weight in 20-metre sized fish schools (see the Supplementary information). This result is consistent with some findings reporting large fish schools up to 1 Km[24].

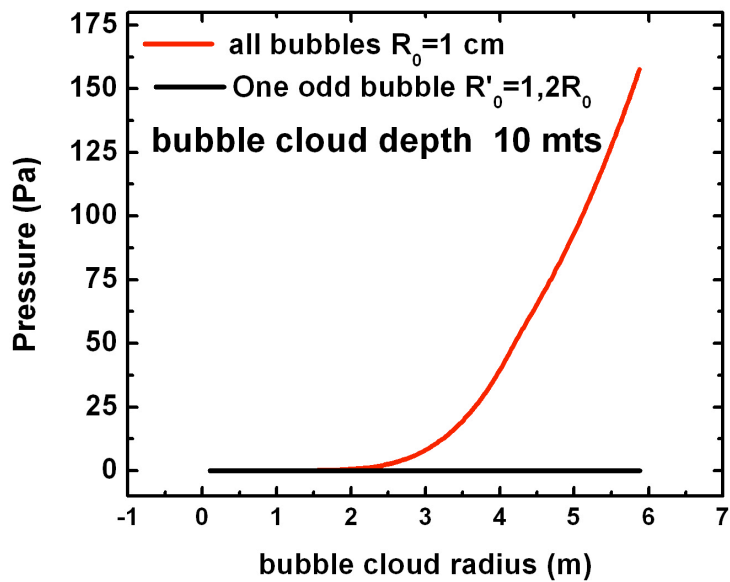

*Figure 3. Total pressure on a bubble at the centre of the cloud as a function of the cloud radius, when either it has the same radius as the rest of the cloud (red curve) or the bubble radius is 20% larger than the rest of the bubbles (black curve). The bubble cloud is located at a depth h=10 metres. We used the same parameters as those in Figure 2 (see text).*

Finally, we would like to discuss the possibility that, through secondary Bjerknes forces, each fish can control at will its inclusion/exclusion to/from the school. Let us assume that a single fish from the school, located at the origin of coordinates (000), changes its gas bladder pressure from $p_0$ to $p'_0$. Then, it will resonate at a different frequency value $\omega'_0$ that is completely detuned from the collective resonance $\omega_1$ of the school fish. Thus, the total Bjerknes force would be

$$F_{SB}^{TOTAL-MS}(000) = \sum_{\forall (ijk \neq 000)} \int_{\Omega_1}^{\Omega_2} F'_{SB}(\omega, ijk, 000) \beta_{ijkmn}(\omega) d\omega \quad (9)$$

where

$$F'_{SB}(\omega, ijk, 000) = \frac{2\pi |A_m|^2}{\varrho_0 \omega^2} \frac{R_0 R'_0}{L_{ijk000}{}^2 |D'_{ijk000}|^2} \left[ \left(\frac{\omega_1^2}{\omega^2} - 1 + \frac{R_0}{L_{ijk000}}\right)\left(\frac{\omega'^2_0}{\omega^2} - 1 + \frac{R'_0}{L_{ijk000}}\right) + \delta'_0 \delta_1 \right] \quad (10)$$

$$|D_{ijkmn}|^2 = \left[ \left(\left(\frac{\omega_1^2}{\omega^2} - 1\right)^2 + \delta_1^2\right)\left(\left(\frac{\omega'^2_0}{\omega^2} - 1\right)^2 + \delta_0'^2\right) - \frac{2 R_0 R'_0}{L_{ijkmn}{}^2}\left(\left(\frac{\omega_1^2}{\omega^2} - 1\right)\left(\frac{\omega'^2_0}{\omega^2} - 1\right) - \delta'_0 \delta_1\right) + \frac{R'_0{}^2 R_0{}^2}{L_{ijkmn}{}^4} \right] \quad (11)$$

The Bjerknes force will be attractive when the sound frequency ω is either ω > $\omega_1$, ω > $\omega_0$' or ω < $\omega_1$, ω < $\omega_0$', and repulsive when $\omega_1$ < ω < $\omega_0$'. Figure 3 shows the total pressure on a bubble located at the centre of the cloud as a function of the cloud radius, either when the bubble has the same radius as other bubbles in the cloud (red curve), or when the bubble radius is 20% larger than the rest of bubbles (black curve). In the second case, the attractive and repulsive forces cancel, releasing the fish from the school.

Finally, this model also explains the short distance repulsive forces that avoid collapsing the whole school. Expression (3) predicts the repulsive forces between neighbouring school mates when they closely approach each other in a similar manner as it appears for the formation of stable bound bubble grapes[18,21,25].

If the BNO is the responsible for the binding forces in fish schools, the experimental evaluation of the PSD magnitude should show strong disturbances (amplification/attenuation) when measured near a fish school, especially at frequency values near the collective resonance $\nu_1$ i.e., the whole fish school would act as a large acoustic antenna near its resonance frequency $\nu_1$, and the BNO would strongly be influenced by the multiple scattering effects of the whole fish school.

**Acknowledgements.** We would like to thank to Prof. C. Feuillade and to Dr. H. Estrada for the critical reading of the manuscript. This work has been partially supported by the Severo